\documentclass[aps,prb,twocolumn,showpacs,10pt]{revtex4-1}
\usepackage{graphicx,amsmath,amssymb,amsfonts,sidecap,color}
\usepackage{empheq}

\newcommand{\ve}{\mathbf}

\begin{document}

\title{RKKY interaction in carbon nanotubes and graphene nanoribbons}

\author{Jelena Klinovaja}
\affiliation{Department of Physics, University of Basel,
             Klingelbergstrasse 82, CH-4056 Basel, Switzerland}
\author{Daniel Loss}
\affiliation{Department of Physics, University of Basel,
             Klingelbergstrasse 82, CH-4056 Basel, Switzerland}

\date{\today}

\begin{abstract}
We study Rudermann-Kittel-Kasuya-Yosida (RKKY) interaction in carbon nanotubes (CNTs) and graphene nanoribbons 
in the presence of spin orbit interactions and magnetic fields. For this we evaluate the static spin susceptibility tensor in real space
in various regimes at zero temperature.
In metallic CNTs
the RKKY interaction depends strongly on the sublattice and, at the Dirac point, is purely ferromagnetic (antiferromagnetic)  for the localized spins on the same (different) sublattice,
whereas in semiconducting CNTs the spin susceptibility depends only weakly on the sublattice and is dominantly ferromagnetic.
The spin orbit interactions break the SU(2) spin symmetry of the system, leading to an anisotropic RKKY interaction of Ising and Moryia-Dzyaloshinsky form, besides the usual isotropic Heisenberg interaction. All these RKKY terms can be made of comparable magnitude by tuning the Fermi level close to the gap induced by the spin orbit interaction.
We further calculate the spin susceptibility also at finite frequencies  and thereby obtain the spin noise in real space via the fluctuation-dissipation theorem.
\end{abstract}



\maketitle


\section{Introduction}

The Rudermann-Kittel-Kasuya-Yosida (RKKY) interaction is an indirect exchange interaction between two localized spins induced by itinerant electrons in a host material.\cite{Ruderman_1954, Kasuya_1956, Yosida_1957}
This effective spin interaction, being determined by the static spin susceptibility, is not only a fundamental characteristics of the host system but also finds interesting and useful applications. One of them is the long-range coupling of spins between distant quantum dots, \cite{Craig_2004, Rikitake_2005} which is needed in scalable quantum computing architectures such as the surface code \cite{Wang_2011} built from spin qubits. \cite{Trifunovic_2012}
In addition, the RKKY interaction, enhanced by electron-electron interactions, can initiate a nuclear spin ordering that leads to striking effects such as helical nuclear magnetism at low temperatures. \cite{Braunecker_2009_PRL, Braunecker_2009_PRB} Such a rotating magnetic field, equivalent to the presence of a uniform magnetic field and Rashba spin orbit interaction (SOI) in one-dimensional systems,\cite{Braunecker_Jap_Klin_2009}  is interesting for Majorana fermion physics in its own right. \cite{Braunecker_Jap_Klin_2009, Flensberg_Rot_Field, Two_field_Klinovaja, klinovaja_Nanoribbon}
The RKKY interaction,
proposed long ago for normal metals of Fermi liquid type,\cite{Ruderman_1954, Kasuya_1956, Yosida_1957,giuliani_book}  was later extended in various ways, in particular to low-dimensional systems with Rashba SOI in the clean \cite{Bruno_2004} and the disordered \cite{stefano_2010}  limit, and to systems with electron-electron interactions in one \cite{Egger_1996,Braunecker_2009_PRL, Braunecker_2009_PRB}
and two dimensions  with \cite{zak_2012} and without \cite{RKKY_2D,Chesi_2008} Rashba SOI.
Also,  a general theorem of Mermin-Wagner type was recently proven for isotropic RKKY systems that excludes magnetic ordering  in one and two dimensions  at any finite temperature but allows it in the presence of SOI. \cite{order_2_2011,Governale_2012}

Moreover, due to  recent progress in  magnetic nanoscale imaging,\cite{amir} one can expect that the direct measurement of the static spin susceptibility has become within experimental reach.\cite{stm_2010}
For this latter purpose, graphene offers the unique advantage over other materials such as GaAs heterostructures in that its surface can be accessed directly on a atomistic scale by the sensing device.
All this makes the spin susceptibility and the RKKY interaction important quantities to study.

Recently, the RKKY interaction in graphene has attracted considerable attention.\cite{Vozmediano_2005,Dugaev_2006, saremi_2007, Black_2010, kogan_2011, Sherafati_2011} Graphene is known for its Dirac-like  spectrum  with a linear dispersion at low energies. This linearity, however, can give rise to  divergences in the expression for the spin susceptibility in momentum space \cite{saremi_2007} and complicates the analysis compared to systems with quadratic dispersion. However, Kogan recently showed that these divergences  can be avoided by working in the Matsubara formalism. \cite{kogan_2011}

In the present work we consider the close relatives of graphene,\cite{dresselhaus_book} namely carbon nanotubes (CNTs) and graphene nanoribbons (GNRs), with a focus on spin orbit interaction and non-uniform magnetic fields.   Metallic CNTs also have a linear spectrum, so for them the imaginary time approach developed in Ref. \onlinecite{kogan_2011} is also most convenient and will be used here. Analogously to graphene, we find that the 
static spin susceptibility changes sign, depending on whether the localized spins belong to the same sublattice or to different sublattices. \cite{footnote_1}
No such dependence occurs for CNTs in the semiconducting regime, characterized by a gap and parabolic spectrum at low fillings. 

The spin orbit interaction in CNTs is strongly enhanced by curvature effects in comparison to flat graphene, \cite{izumida, cnt_helical_2011, klinovaja_cnt} while  in GNRs strong SOI-like effects can be generated by magnetic fields that oscillate or rotate in real space. \cite{Two_field_Klinovaja,klinovaja_Nanoribbon}
Such non-uniform fields can be produced for instance by periodically arranged nanomagnets. \cite{exp_field} 
 Spin orbit effects break the SU(2) spin-symmetry of the itinerant carriers and thus lead, besides the effective Heisenberg interaction, to anisotropic  RKKY terms of Moryia-Dzyaloshinsky  and of Ising form. Quite remarkably, when the Fermi level is tuned close to the gap opened by the SOI, we find that the isotropic and anisotropic terms become of comparable size. This
 has far reaching consequences for ordering in Kondo lattices with RKKY interaction, since this opens up the possibility to have magnetic phase transitions in low-dimensional systems
 at finite temperature that are tunable by electric gates.
 
 We mention that similar anisotropies have been found before for semiconductors with parabolic spectrum and with Rashba SOI in the clean \cite{Bruno_2004} and in the disordered \cite{stefano_2010} limit.  However, the spin orbit interactions in CNTs and in GNRs are of different symmetry and thus both of these problems require a separate study, apart from the fact that the spectrum is linear.

 For all itinerant regimes we consider, the RKKY interaction is found to decay  as $1/R$, where  $R$ is the distance between the localized spins, thus following the standard behavior for RKKY interaction in non-interacting one-dimensional systems. \cite{giuliani_book} [In interacting systems, described by Luttinger liquids, the decay becomes slower. \cite{Egger_1996, Braunecker_2009_PRB,Imambekov_RMF}]
 In contrast, the overall sign as well as the spatial oscillation periods of the RKKY interaction are non-generic and depend strongly on the system and the regimes considered. 
  
 Finally, we will also consider the dynamical spin susceptibility at finite frequency. Via the fluctuation-dissipation theorem we obtain from this the   spin-dependent dynamical structure factor in position space, which describes the  equilibrium correlations of two localized spins separated by a distance $R$.

The paper is organized as follows. Sec. \ref{sec:formalism} contains different approaches to the RKKY interaction including imaginary time formalism for metallic CNTs with a linear spectrum and the retarded Green functions in the real space formalism for semiconducting CNTs. In Sec. \ref{sec:cnts} the low energy spectrum of CNTs is shortly discussed. Afterwards the spin susceptibility is calculated both in the absence of the SOI (Sec. \ref{sec:withou_soi}) and in the presence of the SOI (Sec. \ref{sec:with_soi}). In addition, in Sec. \ref{sec:magnetic} we present results for the case of a magnetic field along the nanotube axis. Such a field breaks both orbital and spin degeneracy, leading to non-trivial dependence of the spin susceptibility on the chemical potential. The fluctuation-dissipation theorem connects the spin susceptibility and the spin fluctuations, allowing us to explore the frequency dependence of the spin noise at zero temperature in Sec. \ref{sec:noise}.
The RKKY interaction in armchair graphene nanoribbons is briefly considered in Sec. \ref{sec:ribbons}. Finally, we conclude with Sec.  \ref{sec:concl} in which we shortly summarize our main results.

\section{Formalism for RKKY \label{sec:formalism}}

The RKKY interaction\cite{Ruderman_1954, Kasuya_1956, Yosida_1957} was studied for a long time and several approaches were developed. In this section we briefly review those used in this work.

The RKKY interaction is an effective exchange interaction between two magnetic spins, $\ve I_i$ and $\ve I_j$, localized at lattice sites $\ve R_i$ and $\ve R_j$, respectively, that are embedded in a system of itinerant electrons with spin-1/2. These electrons have a local spin-interaction  with  the localized spins, described by the Hamiltonian
\begin{equation}
H_{int}= J \sum_{l=i,j} \ve s_l\cdot \ve I_l,
\end{equation}
where $\hbar \ve s_l /2 $ is the electron spin operator at site  $\ve R_l$, and $J$ is the coupling strength.
Using second order perturbation expansion in $J$, \cite{giuliani_book,Bruno_2004,Braunecker_2009_PRB,kogan_2011}
the  RKKY Hamiltonian \cite{Ruderman_1954, Kasuya_1956, Yosida_1957} becomes
\begin{equation}
H_{RKKY} = J^2 \chi^{ij}_{\alpha \beta}  I^\alpha_i  I^\beta_j,
\label{RKKY_general}
\end{equation}
where $\chi^{ij}_{\alpha \beta}=\chi_{\alpha \beta}(\ve R_{ij}, \omega=0)$ is the static (zero-frequency)  spin susceptibility  tensor, and where summation is implied over repeated spin indices $\alpha,\beta=x,y,z$ (but not over $i,j$).
Here, we assumed that the system is translationally invariant so that the susceptibility depends only on the relative distance $\ve R_{ij}\equiv\ve R_{i}-\ve R_{j}$. 
The RKKY interaction can be expressed
in several equivalent ways.
For example, in terms of the retarded Green function $G(\ve R_{ij}; \epsilon+i 0^+)$ the RKKY Hamiltonian is given by
\begin{align}
&H_{RKKY}=-\frac{J^2}{\pi}{\rm Im}  \int_{-\infty}^{\epsilon_F}d\epsilon \ {\rm Tr}\left[(\ve I_i \cdot \ve s) (\ve I_j \cdot \ve s) \right.\nonumber\\
&\ \ \ \ \ \ \ \ \ \ \ \ \ \ \ \  \left. \times G (\ve R_{ij}; \epsilon+i 0^+) G (-\ve R_{ij}; \epsilon+i 0^+)\right],
\label{chi_green}
\end{align}
where the integration over energy $\epsilon$ is limited by the Fermi energy $\epsilon_F$ (see Ref. \onlinecite{Bruno_2004}). Here $\ve s$ is a vector of the Pauli matrices acting on the spin of the itinerant electrons, and the trace Tr runs over the electron spin. The retarded Green function $G (\ve R_{ij}; \epsilon+i 0^+)$, which are spin-dependent here and represented as $2\times 2$-matrices in spin space, are taken in  real ($\ve R_{ij}$) and energy space ($\epsilon$).  
In the presence of spin orbit interaction, we will use Eq. (\ref{chi_green}) as a starting point.

In the absence of spin orbit interaction, the spin is a good quantum number, so
the effective Hamiltonian $H_{RKKY}$ can be significantly simplified, $\chi^{ij}_{\alpha \beta} = \delta_{\alpha\beta}\chi^{ij}_0$, and the RKKY interaction is 
of Heisenberg type (isotropic in spin space).
Expressing the Green functions in terms of the eigenfunctions $\psi_n(\ve R_{i})$ of the electron Hamiltonian, we obtain
\begin{align}
\chi^{ij}_0=2\sum_{n,m} \psi_{n}^*(\ve R_{i}) \psi_{m}(\ve R_{i})&\psi_{n}(\ve R_{j})\psi_{m}^*(\ve R_{j})\nonumber\\
&\times\frac{n_F(\xi_n)-n_F(\xi_m)}{\xi_n - \xi_m},
\label{chi_0}
\end{align}
where the sum runs over all eigenstates of the spinless system, and the factor $2$ accounts for the spin degeneracy. The energy is calculated from the Fermi level, $\xi_n = \epsilon_n - \epsilon_F $, and the Fermi distribution function at $T=0$  is given by $n_F(\xi)=\theta (-\xi)$.

The sum in Eq. (\ref{chi_0}) is divergent in case of a linear spectrum.\cite{yafet_1987, kogan_2011} To avoid these divergences we follow Ref. \onlinecite{kogan_2011} and
work in the imaginary time formalism, again neglecting  the spin structure of the Green functions.
The static real space spin susceptibility at zero temperature is given by 
\begin{equation}
\chi^{ij}_0 = \frac{2}{\hbar} \int_0^{\infty}d\tau\ G_0(\ve R_{ij}, \tau)G_0(-\ve R_{ij}, -\tau),
\label{kogan_chi}
\end{equation}
where the factor 2 again accounts for the spin degeneracy. 
The Matsubara Green functions for $\tau\geq0$ are found as
\begin{align}
G_0(\ve R_{ij}, \pm \tau) = \mp \sum_n& \psi_n^*(\ve R_{i}) \psi_n(\ve R_{j}) e^{\mp \xi_n \tau/\hbar} \theta (\pm \xi_n).
\label{kogan_G}
\end{align}
All three approaches to the RKKY interaction described above [see Eqs. (\ref{chi_green}), (\ref{chi_0}), and (\ref{kogan_chi})] are equivalent to each other.
Which one is used for a particular case depends on calculational convenience.

\section{Carbon nanotubes \label{sec:cnts}}

In this section, we discuss the effective Hamiltonian for a carbon nanotube.
A carbon nanotube is a rolled-up sheet of graphene, a honeycomb lattice  composed of two types of non-equivalent atoms $A$ and $B$. The $(N_1, N_2)$-CNTs can be alternatively  characterized by the chiral angle $\theta$ and the diameter $d$.\cite{dresselhaus_book}
The low-energy physics takes place in two valleys $\bf K$ and $\bf K^\prime$. These two Dirac points are determined by $\ve K = -\ve K^\prime=4\pi ({\bf\hat{t}} \cos \theta + {\bf\hat{z}} \sin \theta)/3a$, where $a$ is the lattice constant. The unit vector ${\bf\hat{z}}$ points along the CNT axis, and ${\bf\hat{t}}$ is the unit vector in the transverse direction. 

\subsection{Effective Hamiltonian}
In the absence of spin orbit interaction CNTs are described by the effective Hamiltonian $H_0$,
\begin{equation}
H_0  = \hbar v_F (k_G\sigma_1 + k \gamma  \sigma_2).
\label{h_0}
\end{equation}
The Pauli matrices $\sigma_i$  act in the space defined by the sublattices $A$ and $B$. The Fermi velocity in graphene $\upsilon_F$ is equal to $10^6\ \rm{m/s}$.
Here,  $\gamma=1\ (\gamma=-1)$ labels the $\ve K$ ($\ve K'$) Dirac points,
and $k$ is the momentum along the $z$-axis calculated from the corresponding Dirac point. The momentum in the circumferential direction $k_G$ is quantized, $k_G=2(m-\gamma\delta/3)/d$, with $d$ the CNT diameter, leading to two kinds of nanotubes: metallic and semiconducting. Here, $m\in \mathbb Z$ is the subband index and $\delta=(N_1-N_2)\ {\rm mod}\,3$ for a $(N_1,N_2)$-CNT (see Ref. \onlinecite{dresselhaus_book}). The spectrum of metallic CNTs (with $k_G=0$) is a Dirac cone, {\it i.e.} linear and  gapless. In contrast to that, the spectrum of semiconducting CNTs (with $k_G\neq 0$) has a gap given by $2\hbar v_F| k_G|$.
In the following we consider only the lowest subband with  
energies 
\begin{equation}
\epsilon_{n}= \nu \hbar v_F \sqrt{k^2+k_G^2},
\label{spect_without_soi}
\end{equation}
where $\nu=1$ ($\nu=-1$) corresponds to electrons (holes), and $n=(k,\gamma,\nu)$ labels the eigenstates.
The corresponding wavefunctions with sublattice spinor are given by
\begin{align}
&\psi_n(\ve R_{i}) =  e^{i (\gamma \ve K+\ve k)\cdot \ve R_{i} } \frac{1}{\sqrt{2}}\begin{pmatrix}
           1 \\ \nu \gamma  e^{i \phi_k}
         \end{pmatrix},\label{functions}\\ 
&e^{i \phi_k} =\frac{k_G+i k}{\sqrt{k_G^2+k^2}},\label{phase}
\end{align}
where $\ve k=(k_G, k)$.
 From now on we redefine Dirac points by shifting the circumferential value of $\ve K$ ($\ve K^\prime$) by $k_G$, so that  $\ve K = -\ve K^\prime=4\pi ({\bf\hat{t}} \cos \theta + {\bf\hat{z}} \sin \theta)/3a+ {\bf\hat{t}} k_G$ and $\ve k = (0, k)$.

\subsection{Spin orbit interaction}
Spin orbit interaction in nanotubes arises mostly from curvature effects, which substantially increase its value in comparison with flat graphene. \cite{izumida, cnt_helical_2011, klinovaja_cnt, cnt_ext_kuemmeth, cnt_ext_kop, cnt_ext_delft}
The effective Hamiltonian, which includes the spin orbit interaction terms,  is given by
\begin{equation}
H_{so}=H_0+\alpha \sigma_1 s_z + \gamma \beta s_z,
\end{equation}
where $s_i$ are the Pauli matrices acting on the spin. The SOI is described by two parameters, $\alpha$ and $\beta$, which depend on the diameter $d$. The values of these parameters can be found in  the framework of the  tight-binding model,  $\alpha = -0.16\ {\rm meV}/d\ {\rm [nm]}$ and $\beta =  -0.62\ {\rm meV} \cos (3\theta) /d\ {\rm [nm]}$ (see Refs. \onlinecite{cnt_helical_2011, klinovaja_cnt}).
The valley index $\gamma$ and the spin projection on the nanotube axis
$s$ are good quantum numbers due to the rotation invariance of the CNT. The conduction band spectrum ($\nu=1$) is given by
\begin{equation}
\epsilon_{n}=\pm\gamma\beta s  + \sqrt{(\hbar \upsilon_F k)^2 + (\gamma \hbar \upsilon_F   k_G + \alpha s)^2 }\, ,
\end{equation}
and the corresponding wavefunctions are given by
\begin{align}
&\psi_{n}(\ve R_{i} ) = e^{i (\gamma \ve K+\ve k)\cdot \ve R_{i} } \frac{1}{\sqrt{2}}\begin{pmatrix}
           1 \\   e^{i  \phi_{s,\gamma}}
         \end{pmatrix} \left| s\right\rangle,\\ 
&e^{i \phi_{s,\gamma}} =\frac{\gamma k_G+s \alpha +i \gamma k}{\sqrt{k^2+(\gamma k_G+ s \alpha )^2}},
\end{align}
where the index $n=(k,\gamma,s)$ labels the eigenstates.
Here $\left| s\right\rangle$, the eigenstate of the Pauli matrix $s_z$, corresponds to the spin state with spin up ($s=1$) or down ($s=-1$). We note that the SOI lifts the spin degeneracy and opens gaps
at zero momentum, $k=0$. In case of  semiconducting nanotubes ($k_G\gg k$), we use the parabolic approximation of the spectrum,
\begin{equation}
\epsilon_{n}=\hbar \upsilon_F k_G + \gamma s (\beta+\alpha)+\hbar \upsilon_F k^2/2 k_G. 
\label{spectrum_simplified}
\end{equation}
Further we  denote the sum of the SOI parameters $\alpha$ and $\beta$ as $\beta_+\equiv\beta+\alpha$. 
Such kind of a spectrum is similar to the spectrum of a CNT in the presence of a pseudo-magnetic field that has opposite signs at opposite valleys. We note that there is a principal difference between a semiconducting CNT and a semiconducting nanowire with Rashba SOI. In the latter, the Rashba SOI can be gauged away by a spin-dependent unitary transformation.\cite{Bruno_2004, Braunecker_Jap_Klin_2009} 
In contrast, the spectrum of CNTs consists of parabolas shifted along the energy axis and not along the momentum axis as in the case of semiconducting nanowires, so the SOI cannot be gauged away. As shown below, this leads to a less transparent dependence of the spin susceptibility on the SOI compared to nanowires. \cite{Bruno_2004}

\section{RKKY in the absence of SOI \label{sec:withou_soi}}

In this section we calculate the spin susceptibility neglecting spin orbit interaction, so all states are two-fold degenerate in spin. We can thus consider a spinless system and account for the spin degeneracy just by introducing a factor of 2 in the expressions for the spin susceptibility, see Eqs. (\ref{chi_0}) and (\ref{kogan_chi}).

\subsection{Metallic nanotubes \label{Sec:matallic}}

The spectrum of a metallic nanotube is linear, see Eq. (\ref{spect_without_soi}), with the momentum in the circumferential direction $k_G$ equal to zero, $k_G=0$. As was mentioned above, in this case the integrals over the momentum in Eq. (\ref{chi_0}) are divergent, \cite{kogan_2011} so it is more convenient to work in the imaginary time formalism [see Eq. (\ref{kogan_chi})], where all integrals remain well-behaved. To simplify notations, we denote the distance between the localized spins as $\ve R \equiv \ve R_i - \ve R_j$ and its projection on the CNT axis as $z$.
Using the wavefunctions given by Eq. (\ref{functions}),
we find the Green functions from Eq. (\ref{kogan_G}), where we replaced sums by integrals, $\sum_k \rightarrow (a/2\pi) \int dk$.  The Green functions on the same sublattices are given by
\begin{align}
G_0^{AA}&(\ve R, \tau) = G_0^{BB}(\ve R, \tau) = - \frac{a }{\pi} \cos ( \ve K \cdot \ve R)\nonumber\\
&\times \frac{ v_F \tau \cos (k_F z)-z \sin (k_F z)}{( v_F \tau)^2 + z^2}.\label{G_met_AA}
\end{align}
The Green function on different sublattices is given by
\begin{align}
G_0^{AB}(\ve R, \tau) &= i\   \frac{a}{\pi} \sin ( \ve K \cdot \ve R) \nonumber\\
 &\times \frac{   v_F \tau \sin (k_F z) + z \cos (k_F z)}{( v_F \tau)^2 + z^2}.\label{G_met_AB}
\end{align}
Here, the Fermi wavevector $k_F$ is determined by the Fermi energy $\epsilon_F$  as $k_F=\epsilon_F/ \hbar \upsilon_F$. 
For the corresponding spin susceptibilities [see Eq. (\ref{kogan_chi})] we then obtain after straightforward integration
\begin{align}
\chi_0^{AA}(\ve R)=&\frac{-a^2}{4\pi \hbar v_F |z|}
\left[1+\cos( 2\ve K\cdot \ve R)\right]\cos (2 k_F z),\label{met_aa}\\
\chi_0^{AB}(\ve R)=&\frac{ a^2}{4\pi \hbar v_F |z|}\left[1-\cos( 2\ve K\cdot \ve R)\right]\cos (2 k_F z).
\label{susc_metallic_cnt}
\end{align}

If the chemical potential is tuned strictly to the Dirac point, $k_F=0$, the spin susceptibility is purely ferromagnetic for the atoms belonging to the same sublattices, $\chi_0^{AA}, \chi_0^{BB} \leq 0$, whereas it is purely antiferromagnetic for the atoms belonging to different sublattices, $\chi_0^{AB}\geq 0$. \cite{footnote_1}
For the chemical potential tuned away from the Dirac point we observe in addition to the sign difference oscillations of the spin susceptibility in real space with period of half the Fermi wavelength $\pi/k_F$. This oscillation, together with the $1/z$ decay, is typical for RKKY interaction in one-dimensional systems.\cite{giuliani_book}

\begin{figure}[!tp]
\centering
\includegraphics[width=0.3\textwidth]{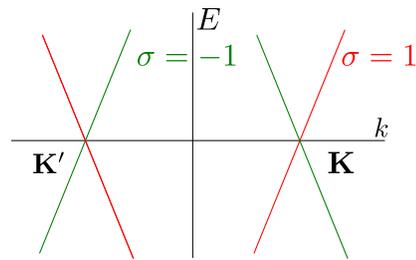}
\caption{The Dirac spectrum of a metallic nanotube. Each branch is characterized by the isospin value $\sigma$, which is an eigenvalue of the Pauli matrix $\sigma_2$. We note that the Kramers partners at $K$ and $K'$, respectively, are characterized by the same value of the isospin, and two partner states at the same cone are characterized by opposite isospins.}
\label{isospin}
\end{figure}

An immediate consequence  of the opposite signs of  the susceptibilities in Eqs. (\ref{susc_metallic_cnt}) is that any ordering of spins localized at the honeycomb lattice sites will be antiferromagnetic. Such order produces a staggered magnetic field that can act back on the electron system and give rise to scattering of electrons between branches of opposite isospin $\sigma$ at the same Dirac cone (see Fig. \ref{isospin}). It has been shown elsewhere that such backaction effects can lead to a spin-dependent Peierls gap in the electron system. \cite{Braunecker_2009_PRL, Braunecker_2009_PRB,Braunecker_Jap_Klin_2009}

\subsection{Semiconducting nanotubes \label{sec:semiconducting}}

\subsubsection{Zero chemical potential }

Now we consider semiconducting CNTs that are characterized by a non-zero circumferential wavevector $k_G$ and a corresponding gap in the spectrum.  We begin with the case of the Fermi level $\epsilon_F$ lying in the middle of the gap, $\epsilon_F=0$. As a result, there are no  states at the Fermi level. This leads to  a strong suppression of the RKKY interaction.  For example, the Green function on the same sublattice, found from Eqs. (\ref{kogan_G})  and (\ref{functions}), is given by
\begin{align}
G^{AA}(\ve R, \tau) \approx -\frac{a}{2\pi} &\cos( \ve K\cdot \ve R)e^{i  k_G z - \upsilon_F |\tau| k_G }\nonumber\\
&\times \sqrt{\frac{2\pi k_G}{ \upsilon_F|\tau|}} e^{ -  z ^2 k_G /2 \upsilon_F |\tau| }, \label{aa_sem_gap_p}
\end{align}
where we used the simplified  parabolic spectrum, see Eq. (\ref{spectrum_simplified}). The spin susceptibility is then obtained from Eq. (\ref{kogan_chi}),
\begin{align}
\chi^{AA}(\ve R)&=-\frac{2a^2 k_G}{\pi \hbar \upsilon_F}\cos^2( \ve K\cdot \ve R)K_0(2\sqrt{2}k_G |z| ),
\end{align}
where $K_0(x)$ is the modified Bessel function of  second kind, which decays exponentially at large distances, $K_0(x)\approx e^{-x}\sqrt{\pi/2x}$ for $x \gg 1$.
This exponential decay (on the scale of the CNT diameter $d$) of the spin susceptibility in the case when the Fermi level lies in the gap is not surprising. There are just no delocalized electron states that can assist the effective coupling between two separated localized spins. From now on we neglect any contributions coming from higher or lower bands.

\subsubsection{Non-zero chemical potential}
In this subsection we assume that the Fermi level is tuned in such a way that it crosses, for example, the conduction band. 
Thus, the itinerant states assist the RKKY interaction between localized spins.  The spectrum of a semiconducting CNT is parabolic [see Eq. (\ref{spectrum_simplified})], so it is more convenient to work with the spin susceptibility 
given by Eq. (\ref{chi_0}). In momentum space the spin susceptibility on the same sublattice is given by $\chi^{AA}(q)=\sum_{\gamma, \gamma'}\chi^{AA}(\gamma, \gamma'; q) $ with
\begin{equation}
\chi^{AA}(\gamma, \gamma'; q)=\frac{1}{2}\sum_{k}\frac{\theta[\xi_{\gamma'}(k+q)]-\theta[\xi_{\gamma}(k)]}{\xi_{\gamma'}(k+q)-\xi_{\gamma}(k)}.
\end{equation}
Performing integration over momentum $k$, we arrive at an expression 
similar to the Lindhard function,
\begin{align}
\chi^{AA}(\gamma, \gamma'; q)=-\frac{ak_G}{ 2 \pi \hbar \upsilon_F q} \ln \left| \frac{q+ 2k_F}{q-2k_F} \right|,
\label{log}
\end{align}
where the Fermi momentum $k_F$ is defined as $k_F =\sqrt{ 2 k_G( \epsilon_F - \hbar \upsilon_F k_G)/ \hbar \upsilon_F}$. Next we go to real space by taking the Fourier transform of $\chi^{AA}(\gamma, \gamma'; q)$. This can be readily done
by closing the integration contour in the upper (lower) complex plane for $z>0$ $(z<0)$ and deforming it around the two branch cuts of the logarithm in Eq.~(\ref{log}) that run from $\pm 2k_F$ to $\pm \infty$. This yields,
\begin{align}
&\chi^{AA}(\ve R) = \frac{a^2k_G}{\pi \hbar \upsilon_F } {\rm si} (2k_F   |z|)\left[1+\cos( 2\ve K\cdot \ve R)\right].
\label{semi_AA}
\end{align}
Here, the sine integral is defined as
\begin{equation}
{\rm si}(x)=\int_0^x dt\ \frac{\sin t}{t}  - \frac{\pi}{2},
\end{equation}
and at large distances, $x\gg1$, its asymptotics is given by ${\rm si}(x)~\approx~-\cos (x)/x$.

The evaluation of the spin susceptibility for different sublattices is more involved,
\begin{equation}
\chi^{AB}(\gamma, \gamma'; q)=\frac{\gamma\gamma'}{2}\sum_{k}e^{-i \Delta \phi_{k,q}}\frac{\theta[\xi_{\gamma'}(k+q)]-\theta[\xi_{\gamma}(k)]}{\xi_{\gamma}(k)-\xi_{\gamma'}(k+q)},
\end{equation}
where the phase difference  given by $\Delta \phi_{k,q}=\phi_k - \phi_{k+q} $ depends on the momenta $k$ and $q$, see Eq. (\ref{phase}). Taking into account that $k_G$ is the largest momentum characterizing the system, we expand the phase factor as
 $e^{i(\phi_k-\phi_{k+q})}\approx 1-i q/k_G$. The main contribution to the spin susceptibility comes from the momentum-independent part and is given by
\begin{align}
\chi^{AB}(\ve R) = \frac{a^2k_G}{\pi \hbar \upsilon_F} {\rm si} (2k_F   |z|)\left[1-\cos( 2\ve K\cdot \ve R)\right].
\label{semi_AB}
\end{align}
In the next step we evaluate the correction $\Delta \chi^{AB}(\ve R)$ to the spin susceptibility $\chi^{AB}(\ve R)$ arising from the momentum-dependent part in the phase factor $e^{i \Delta \phi_{k,q}}$. In momentum space it is given by
\begin{align}
\Delta \chi^{AB}(\gamma, \gamma'; q)=\frac{i a\gamma\gamma'}{2 \pi \hbar \upsilon_F} \ln \left| \frac{q+ 2k_F}{q-2k_F} \right|.
\label{delta_chi_ab}
\end{align} 
By taking the Fourier transform of Eq. (\ref{delta_chi_ab}), we arrive at the following expression,
\begin{align}
\Delta \chi^{AB}(\ve R)=&\sum_{\gamma,\gamma'} e^{i(\gamma-\gamma') \ve K\cdot \ve R  } \frac{ia^2\gamma\gamma'}{ 2\pi \hbar \upsilon_F}\nonumber\\
&\times \int^{\infty}_{-\infty} dq\  e^{i q z}  \ln \left| \frac{q+ 2k_F}{q-2k_F} \right|.
\label{delta_chi}
\end{align}
The integral in Eq. (\ref{delta_chi}) can be evaluated easily  by recognizing it as the derivative of ${\rm si}(x)$ [see Eq.~(\ref{log})], 
\begin{align}
I=\int^\infty_{-\infty} &dx\ e^{i \alpha x} \ln \left| \frac{x+1}{x-1} \right|= -i \frac{\sin \alpha}{|\alpha|}\label{exact_I},
\end{align}
where $\alpha$ is real.
As a result, the correction to the spin susceptibility on different sublattices $\chi^{AB} (\ve R)$ is given by
\begin{equation}
\Delta \chi^{AB} (\ve R)= \frac{a^2\sin (2k_Fz)}{\pi \hbar \upsilon_F |z|}\left[1-\cos( 2\ve K\cdot \ve R)\right].
\end{equation}
We note that $\Delta \chi^{AB} (\ve R)$ is small in comparison with $\chi^{AB} (\ve R)$ by a factor $k_F/k_G\ll1$, and, thus, this correction plays a role only around the points where the oscillating function $\chi^{AB} (\ve R)$ vanishes. We emphasize that the spin susceptibility for semiconducting CNTs does not possess any significant dependence on the sublattices in contrast to metallic CNTs.

\section{RKKY in the presence of SOI \label{sec:with_soi}}

\begin{figure*}[t]
\centering
\includegraphics[width=0.7\textwidth]{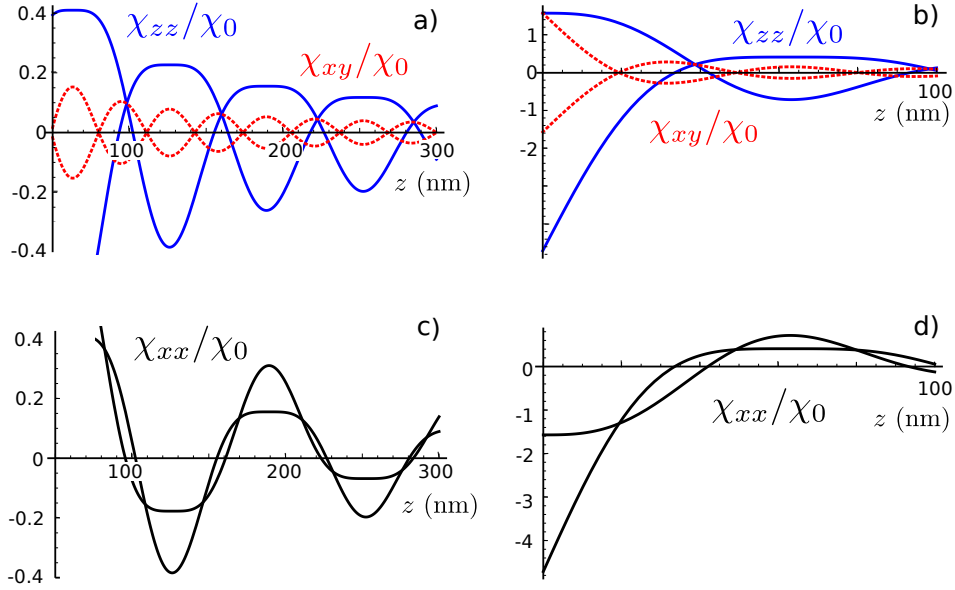}
\caption{The  spin susceptibility $\chi_{\alpha \beta}$, given in Eqs.~(\ref{chi_zz})-(\ref{chi_xz}), plotted as a function of  distance between localized spins $z$, for a semiconducting (11,1)-CNT
in the presence of spin orbit interaction and at zero B-field. Here, $\chi_0=a^2 k_G/ 2\pi \hbar \upsilon_F$.
For clarity we suppress the fast oscillating factors and plot only the slowly varying envelopes at small (a,c) and large scales (b,d). The chemical potential $\mu=472\ \rm meV$ ($\delta \mu \equiv \mu - \hbar \upsilon_F k_G +\beta_+ =1\ \rm meV$) is tuned inside the gap opened by SOI with corresponding value $2\beta_+=1.7\ \rm meV$ for a (11,1)-CNT. \cite{cnt_helical_2011} The diagonal components $\chi_{zz}$ [(a,b) blue full curve] and $\chi_{xx}$ (c,d) oscillate with period $2\pi /k_{+,-}\approx 2\pi /k_F$. In contrast to that, the off-diagonal component $\chi_{xy}$ [(a,b) red dashed curve], oscillates with period $2\pi /k_{+,+}\approx \pi / k_F$. All components decay as $1/z$. Note that  the diagonal and off-diagonal components are of comparable size.}
\label{plot_chi1}
\end{figure*}

In the presence of spin orbit interaction, the spin space is no longer invariant under rotations, and as a consequence the spin susceptibility is described by the tensor $\chi_{\alpha \beta}^{ij}$   [see Eq. (\ref{RKKY_general})] with non-vanishing off-diagonal components.  
In this case it is more convenient to work in the framework of retarded Green functions\cite{Bruno_2004}
in which the RKKY Hamiltonian $H_{RKKY}$ is
given by Eq. (\ref{chi_green}). Below we neglect the weak dependence of the susceptibility on the sublattice discussed above and focus on the SOI effects.
The Green functions in the energy-momentum space can be expressed as
\begin{align}
G(k, \epsilon+i 0^+)=\sum_\gamma [G_0(\gamma, k,\epsilon)  + G_1(\gamma, k,\epsilon)  s_z],
\label{eq_g_k}
\end{align}
where the diagonal and off-diagonal (in spin space) Green functions are given by
\begin{align}
&G_0(\gamma, k,\epsilon) =\frac{k_G}{\hbar \upsilon_F}\sum_{\eta=\pm 1} \frac{1}{k_{ \gamma\eta}^2- k^2 + i 0^+},\\
&G_1(\gamma, k,\epsilon) =-\frac{k_G}{\hbar \upsilon_F}\sum_{\eta=\pm 1} \frac{\eta}{k_{\gamma\eta}^2- k^2 + i 0^+}.
\end{align}
Here, to simplify notations, we introduced the wavevector $ k_{\eta'}$, 
defined as (with ${\eta'}={ \gamma \eta}$)
\begin{equation}
k_{\eta'} = \sqrt{2k_G (\epsilon - \hbar \upsilon_F k_G  - \eta' \beta_+)/\hbar \upsilon_F},
\end{equation}
which can take both real and imaginary values.
In a next step we transform the Green functions from momentum to real space,
\begin{equation}
G_{0,1}(\ve R, \epsilon+i 0^+)=\frac{a}{2}\sum_\gamma\int \frac{dk}{2\pi} G_{0,1}(\gamma, k,\epsilon) e^{i  (\gamma\ve K+\ve k)\cdot \ve R},
\end{equation}
leading to
\begin{align}
G_0(\ve R, \epsilon)&=-i \frac{ak_G}{2\hbar \upsilon_F }\cos (\ve K\cdot \ve R) \sum_{\eta=\pm1}\frac{e^{i k_{\eta} |z|}}{k_{\eta}+ i 0^+} \label{g0} ,\\
G_1(\ve R, \epsilon)&=\frac{ak_G}{2\hbar \upsilon_F }\sin (\ve K \cdot \ve R)\sum_{\eta=\pm1}\frac{\eta e^{i k_{\eta} |z|}}{k_{\eta}+ i 0^+}.\label{g1}
\end{align}

Substituting $G(\ve R, \epsilon) = G_0(\ve R, \epsilon) + G_1(\ve R, \epsilon) s_z$ into Eq. (\ref{chi_green}), we find for the RKKY Hamiltonian,
\begin{align}
H&_{RKKY}=-\frac{ J^2 }{\pi}{\rm Im} \left[ -4 I_i^z I_j^z \int_{-\infty}^{\epsilon_F} d\epsilon\ G^2_1(\ve R, \epsilon)
\right.\nonumber\\
&\left.-4i (\ve I_i\times \ve I_j)_z\int_{-\infty}^{\epsilon_F} d\epsilon\ G_1(\ve R, \epsilon)G_0(\ve R, \epsilon)\right.\nonumber\\
&\left.+2\ve I_i\cdot \ve I_j\int_{-\infty}^{\epsilon_F} d\epsilon\ \left[G^2_0(\ve R, \epsilon)+G^2_1(\ve R, \epsilon)\right]\right].\label{Hrkky_g}
\end{align}
Here, the trace over spin degrees of freedom were calculated by using the following identities
\begin{align}
 & {\rm Tr}\{(\ve I_i\cdot \boldsymbol s) (\ve I_j \cdot \boldsymbol s)\}=2\ve I_i\cdot \ve I_j,\\
&{\rm Tr}\{(\ve I_i\cdot \boldsymbol s)s_z(\ve I_j \cdot \boldsymbol s)\}
=-2i (\ve I_i\times \ve I_j)_z,\\
&{\rm Tr}\{(\ve I_i\cdot \boldsymbol s) s_z(\ve I_j \cdot \boldsymbol s)s_z\}=2(2  I_i^z  I_j^z - \ve I_i\cdot \ve I_j ).
\end{align} 
All integrals in Eq. (\ref{Hrkky_g}) are of the same type,
\begin{align}
{\rm Im} \int_{-\infty}^{\epsilon_F} d\epsilon\ &\frac{e^{i  k_{\eta} |z|}}{k_{\eta}+i 0^+}\frac{e^{i  k_{\eta'} |z|}}{k_{\eta'}+i 0^+}
=\frac{\hbar \upsilon_F}{k_G}{\rm si}( k_{\eta \eta'} |z|),
\end{align}
and can be easily evaluated by changing variables from  the original $\epsilon$ to $k_{\eta} + k_{\eta'}$ . We denote the real part of the sum of two Fermi wavevectors 
as $k_{\eta, \eta'}= {\rm Re}[k_{\eta}(\epsilon_F)+k_{\eta'}(\epsilon_F)]$ with the indices $\eta, \eta' = \pm 1$. As a result, we arrive at
the   RKKY Hamiltonian in the form of Eq. (\ref{RKKY_general}), 
where the components of the spin susceptibility tensor $\chi^{ij}_{\alpha \beta}\equiv\chi_{\alpha \beta} (\ve R)$ are explicitly given by 
\begin{align}
&\chi_{zz}=\frac{a^2k_G}{2\pi\hbar \upsilon_F }\Big[{\rm si}( k_{+,+}  |z|)+{\rm si}( k_{-,-} |z|)\nonumber\\
&\ \ \ \ \ \ \ \ \ \ \ \ \ \ \ \ \ \  +2 \cos (2\ve K\cdot \ve R)  {\rm si} (k_{-,+}|z|)\Big],\label{chi_zz}\\
&\chi_{xx}=\frac{a^2k_G}{2\pi\hbar \upsilon_F }\Big(2  {\rm si} (k_{-,+}|z|) +\cos (2\ve K\cdot \ve R)\nonumber\\
&\ \ \ \ \ \ \ \ \ \ \ \ \ \ \ \ \ \ \ \ \times[{\rm si}( k_{+,+}  |z|)+{\rm si}( k_{-,-} |z|)]\Big),\\
&\chi_{xy}=\frac{a^2k_G}{2\pi\hbar \upsilon_F }\sin (2\ve K\cdot \ve R)\nonumber\\
&\ \ \ \ \ \ \ \ \ \ \ \ \ \ \ \ \ \ \ \ \times\Big[ {\rm si}(k_{+,+}  |z|)-{\rm si}(k_{-,-} |z|)\Big],
\label{chi_xz}
\end{align}
with $\chi_{xx}=\chi_{yy}$, $\chi_{xy}=-\chi_{yx}$, and all other components being zero.
First, we note that the off-diagonal components $\chi_{xy}, \chi_{yx}$ are non-zero. They describe the response to a perturbation applied perpendicular to the 
nanotube axis.
This opens up the possibility to test the presence of SOI in the system by measuring off-diagonal components of the spin susceptibility tensor $\chi_{\alpha \beta}$.
Second, the spin response in a direction perpendicular to the $z$-axis cannot be caused by a perturbation along the $z$-axis,  thus $\chi_{yz}=\chi_{xz}=0$. This simply reflects the rotation-invariance of CNTs around their axes. The difference between the diagonal elements of the spin susceptibility tensor, $\chi_{xx}=\chi_{yy}$ and $\chi_{zz}$, again arises from the SOI and is another manifestation of the broken rotation invariance of spin space.
In total this means that the RKKY interaction given in Eq. (\ref{Hrkky_g}) is anisotropic in the presence of SOI, giving rise 
to an Ising term $\propto I_i^z  I_j^z$ and a Moryia-Dzyaloshinsky term
$\propto (\ve I_i\times \ve I_j)_z$, in addition to the isotropic Heisenberg term $\propto \ve I_i\cdot \ve I_j$.

\begin{figure}[!bt]
\centering
\includegraphics[width=0.45\textwidth]{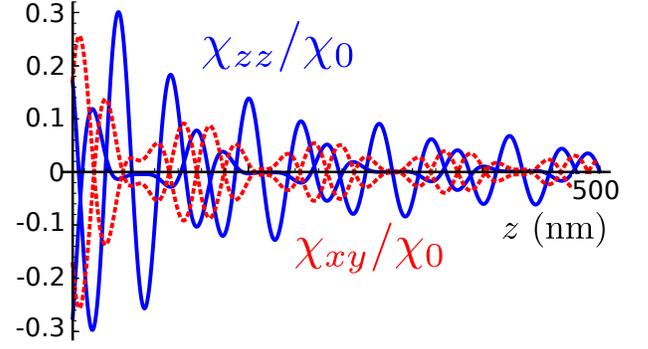}
\caption{The same as Fig.~\ref{plot_chi1} (a,b) but with the chemical potential $\mu=474\ \rm meV$ being tuned above the gap 
opened by SOI. The spin susceptibility decays as $1/z$ and exhibits beatings, with period determined by the SOI parameters.}
\label{beatings}
\end{figure}

Quite remarkably, when the Fermi level is tuned close to the gap opened by the SOI, then the off-diagonal and diagonal components of the susceptibility tensor become of comparable 
magnitude, see Figs. \ref{plot_chi1} and \ref{beatings}. This has important consequences for a Kondo lattice system, where a highly anisotropic RKKY interaction will give rise to an ordered magnetic phase even at finite temperatures.\cite{order_2_2011} 
As a potential candidate for such a Kondo lattice \cite{Braunecker_2009_PRB} we might mention
a CNT made out of the  $^{13}C$-isotope, \cite{Churchill_2009} where each site of the graphene lattice contains a nuclear spin-1/2 to which the itinerant electrons couple via hyperfine interaction. \cite{Fischer_2009}

We note that the susceptibility depends on two Fermi wavevectors via $k_{\eta \eta'}$
in a rather complicated way (see Figs. \ref{plot_chi1} and \ref{beatings}).
In the absence of  SOI, we recover the result for the spin susceptibility on the same sublattice,
$\chi^{AA}(\ve R)$ [see Eq. (\ref{semi_AA})].
The leading term in the spin susceptibility for different sublattices, $\chi^{AB}(\ve R)$, can be obtained from Eqs. (\ref{chi_zz}) - (\ref{chi_xz}) by putting a minus sign in front of $\cos (2\ve K\cdot \ve R)$. In addition, as shown in Sec. \ref{sec:semiconducting}, the spin susceptibility vanishes if the Fermi level is tuned inside the gap in semiconducting CNTs, so that both $k_\pm(\epsilon_F)$ are purely imaginary.
 If the chemical potential is inside the gap opened by the SOI, the Fermi wavevector $k_-(\epsilon_F)$ is still imaginary, at the same time $k_+(\epsilon_F)$ is real, giving $k_{+,+}=2k_+(\epsilon_F)$, $k_{-,-}=0$, and $k_{+,-}=k_+(\epsilon_F)$. This results in the behavior  of the  spin susceptibility shown in Fig. \ref{plot_chi1}. The strength of the RKKY interaction decays oscillating as $1/R$. The oscillation period is determined by $k_{+,+}$ for $\chi_{xy}$ and by $k_{+,-}$ for $\chi_{xx}$ and $\chi_{zz}$, see Fig. \ref{plot_chi1}. If the chemical potential is above the SOI gap, then both wavevectors $k_\pm(\epsilon_F)$ are real, giving rise to oscillations with two different frequencies that result in beating patterns for the spin susceptibility, see Fig. \ref{beatings}. 

\section{RKKY with magnetic field \label{sec:magnetic}}

In Sec. \ref{sec:with_soi} we demonstrated that the presence of SOI, which breaks rotation invariance in spin space, leads to an anisotropic spin susceptibility.
Another way to break this rotation invariance is to  apply a magnetic field, which also breaks  time-reversal invariance. In this section we again neglect sublattice asymmetries discussed above and focus on the effects of a magnetic field $\bf B$ applied along the nanotube axis $z$ for a semiconducting nanotube.  The Zeeman term $H_Z=\Delta_Z s_z=\mu_B B s_z$ lifts the spin degeneracy. Here, $\mu_B$ is the Bohr magneton. The orbital term $H_{orb}=\gamma \hbar \upsilon_F k_{mag}$ leads to a shift of the transverse wavevector $k_G$ by $k_{mag}$, which finds its origin in the Aharonov-Bohm effect and is given by   $k_{mag} = \pi B d |e|/2 h$ for a nanotube of diameter $d$. Thus, the valley degeneracy of the levels is also lifted. The spectrum of the effective Hamiltonian $H_0+H_Z+H_{orb}$ in the case of a semiconducting nanotube is given by 
\begin{equation}
\epsilon_{\gamma, s}=\hbar \upsilon_F k_{G,\gamma} + s (\gamma\beta_++\Delta_z)  +\hbar \upsilon_F k^2/2 k_{G,\gamma},
\label{xi_magnetic}
\end{equation}
where $k_{G,\gamma}=k_G +\gamma k_{mag}$.
The Green functions in momentum space can be found similar to Eq. (\ref{eq_g_k}). As a result,
we arrive at the following expression for the Green functions,
\begin{align}
G(k, \epsilon+i 0^+)=\sum_\gamma [G_0(\gamma, k,\epsilon) + G_1(\gamma, k,\epsilon)  s_z],
\end{align}
where
\begin{align}
&G_0(\gamma, k,\epsilon) =\frac{k_{G,\gamma}}{\hbar \upsilon_F}\sum_s \frac{1}{\kappa_{\gamma, s}^2- k^2 + i 0^+},\\
&G_1(\gamma, k,\epsilon) =-\frac{k_{G,\gamma}}{\hbar \upsilon_F}\sum_s \frac{ s}{\kappa_{\gamma, s}^2- k^2 + i 0^+}.
\end{align}
Here, we define wavevectors $\kappa_{\gamma, s}$ as a function of the energy $\epsilon$ from Eq. (\ref{xi_magnetic}) as
\begin{equation}
 \kappa_{\gamma, s} = \sqrt{\frac{2k_{G,\gamma} [\epsilon - \hbar \upsilon_F k_{G,\gamma} - s (\gamma\beta_++\Delta_z)]}{\hbar \upsilon_F}},
\end{equation}
which can take both non-negative real and imaginary values. 
 The Green functions in real space are found by Fourier transformation,
\begin{align}
G_0(\ve R, \epsilon)&=-i \sum_{\gamma,s}\frac{ak_{G,\gamma}}{2\hbar \upsilon_F }e^{i\gamma \ve K\cdot \ve R} \frac{e^{i \kappa_{\gamma, s} |z|}}{\kappa_{\gamma, s}+ i 0^+} \label{g0B} ,\\
G_1( \ve R, \epsilon)&=-i \sum_{\gamma,s}\frac{ak_{G,\gamma}}{2\hbar \upsilon_F }e^{i\gamma \ve K\cdot \ve R} \frac{ s e^{i  \kappa_{\gamma, s} |z|}}{ \kappa_{\gamma, s}+ i 0^+}.\label{g1B}
\end{align}
By substituting Eqs. (\ref{g0B}) and (\ref{g1B}) into Eq. (\ref{Hrkky_g}), 
we arrive at the effective RKKY Hamiltonian. Since $ k_{mag}/k_G \ll 1$, we can neglect the dependence of the spectrum slope on the magnetic field, which simplifies the calculations considerably.

\begin{figure}[!bt]
\centering
\includegraphics[width=0.45\textwidth]{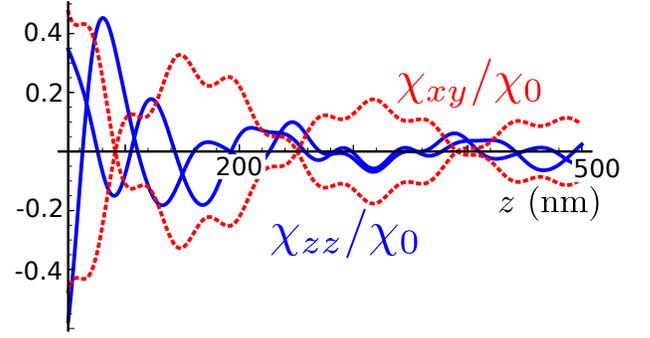}
\caption{The same as Fig.~\ref{plot_chi1} (a,b) but with the chemical potential $\mu=472.7\ \rm meV$ being tuned in such a way that there are three pairs of states at the Fermi level and with a magnetic field $B=1\ {\rm T}$. The spin susceptibility decays as $1/z$ and exhibits beatings, with period determined by the SOI parameters and the magnetic field.}
\label{fig:magnetic}
\end{figure}

 At the end we arrive at the following expressions for the spin susceptibility tensor components
 \begin{align}
&\chi_{xx}=\frac{a^2k_{G}}{4\pi\hbar \upsilon_F }\nonumber\\
&\times \sum_{\gamma, s}\Big[{\rm si}(k_{\gamma,s; \gamma,\bar{s}}|z|)+\cos(2 \ve K\cdot \ve R){\rm si}(k_{\gamma,s; \bar{\gamma},\bar{s}}|z|)\Big],\\
&\chi_{zz}=\frac{a^2k_{G}}{4\pi\hbar \upsilon_F }\nonumber\\
&\times \sum_{\gamma, s}\Big[{\rm si}(k_{\gamma,s; \gamma,s}|z|)+\cos[2 \ve K\cdot \ve R]{\rm si}(k_{\gamma,s; \bar{\gamma},s}|z|)\Big],\\
&\chi_{xy}=\frac{a^2k_{G}}{4\pi\hbar \upsilon_F } \sum_{\gamma, s}\gamma s \sin[2 \ve K\cdot \ve R]{\rm si}(k_{\gamma, s; \bar{\gamma}, \bar{s}}|z|),
\label{chi_fieldxy}
\end{align}
with $\chi_{xx}=\chi_{yy}$, $\chi_{xy}=-\chi_{yx}$, and the rest being zero. Here, we use the notation $k_{\gamma, s; \gamma', s'}={\rm Re}[\kappa_{\gamma, s}(\epsilon_F)+\kappa_{ \gamma', s'}(\epsilon_F)]$. Again, the RKKY interaction decays at large distances as $1/R$.
The spin susceptibility also exhibits oscillations and beating patterns (similar to ones shown in Fig. \ref{beatings}) determined by four different Fermi wavevectors $k_{\gamma s}(\epsilon_F)$, see Fig. \ref{fig:magnetic}. 
Finally, we note that the same beating patterns arises also for nanowires with parabolic spectrum in the presence of a magnetic field giving rise to a Zeeman splitting.

\section{Spin Fluctuations \label{sec:noise}}

The spin susceptibility is a fundamental characteristics of the system. At zero frequency it describes the RKKY interaction between localized spins. At finite frequencies $\omega$ the spin susceptibility gives access to the equilibrium spin noise in the system.
For a general observable, $s_\alpha$, the fluctuation-dissipation theorem\cite{giuliani_book} 
 connects the dynamical structure factor  $S_{\alpha \alpha}({\bf q},\omega)=(1/2\pi) \int dt\ \langle s_{\alpha}({\bf q},t)s_{\alpha}({\bf -q},0)\rangle e^{i\omega t}$, describing equilibrium fluctuations, with the linear response susceptibility,  $\chi_{\alpha \alpha} ({\bf q},\omega)=(-i/\hbar) \int dt\  \Theta(t) \langle[s_{\alpha}({\bf q},t),s_{\alpha}({\bf -q},0)]\rangle e^{i\omega t}$,
\begin{equation}
{\rm Im} \chi_{\alpha \alpha} ({\bf q},\omega)  =\frac{\pi}{ \hbar}(e^{-\frac{\hbar \omega}{ k_B T}} -1)S_{\alpha \alpha}({\bf q},\omega)\, , 
\label{FDT}
\end{equation}
for $\omega >0$.
Below we calculate the spin susceptibility for both metallic and semiconducting nanotubes at finite frequencies, and  obtain via Eq. (\ref{FDT}) the spin correlation function (spin noise) $S_{\alpha \alpha}(\ve R, \omega)$ in real space at zero temperature,
\begin{equation}
S_{\alpha \alpha} (\ve R, \omega) = -\frac{\hbar}{\pi}\ {\rm Im} \chi_{\alpha \alpha} ({\bf R},\omega),
\label{zero_temperature}
\end{equation}
Here, we assumed that the system is invariant under parity transformation so that $\chi({\bf q},\omega)$ becomes an even function of ${\bf q}$.
For simplicity, we consider only the  case without SOI.

\subsection{Metallic nanotubes}
The spin susceptibility at finite Matsubara frequencies $\omega_n$
is given by
\begin{equation}
\chi_0(\ve R, i\omega_n) = \frac{2}{{\hbar}} \int_0^{\infty}d\tau\ G_0(\ve R, \tau)G_0(-\ve R, -\tau) e^{i\omega_n \tau},
\label{kogan_chi_freq}
\end{equation}
where we modified  Eq. (\ref{kogan_chi}) accordingly, and the Green function $G_0(\ve R, \tau)$ was found before [see Eqs. (\ref{G_met_AA}) and (\ref{G_met_AB})].
For the spin susceptibility on the same sublattice 
we get
\begin{align}
&\chi^{AA}_0( \ve R, i\omega_n)=\frac{2a^2}{\pi^2\hbar}\cos^2 ( \ve K \cdot \ve R)\nonumber \\
&\times\int_0^\infty d\tau\ e^{i\omega_n \tau}\frac{[z \sin (k_F z)]^2-[v_F \tau \cos (k_F z)]^2}{[( v_F \tau)^2 + z^2]^2}.
\end{align}
Introducing the notation
\begin{align}
&I_n(x)=\int_0^\infty d\tau\  \frac{\tau^n e^{i  \tau x}}{(1+\tau^2)^2}\, ,
\end{align}
where $n=0,2$, the susceptibility can be rewritten as
\begin{align}
&\chi_0^{AA}(\ve R, i\omega_n)=\frac{2a^2}{\pi^2\hbar v_F |z|}\cos^2 ( \ve K \cdot \ve R)\nonumber\\
&\times \left[ \sin^2 (k_F z) I_0 \left(\frac{\omega_n |z|}{  v_F}\right) - \cos^2 (k_F z) I_2 \left(\frac{\omega_n |z|}{ v_F}\right)
\right].
\end{align}
The asymptotics of $I_n(x)$ for small $x$ is given by $I_0(x)\approx \pi/4+ i x/2$ and $I_2(x) \approx \pi(1-2x)/4 - i x (\ln x + \gamma_e -1/2)$, where $\gamma_e$ is the Euler constant. Performing the standard analytic continuation from Matsubara to real frequencies, \cite{giuliani_book} we obtain the spin susceptibility  $\chi^{AA}_0(\omega)$, and from its imaginary part  the spin noise, see Eq. (\ref{zero_temperature}).  Explicitly, the dynamical structure factor at zero temperature and for $ { v_F}\gg {\omega |z|}\geq 0$ is given by
\begin{align}
S^{AA}(\ve R,\omega)=-
\frac{2a^2\omega}{\pi^2 v_F^2}\cos^2 ( \ve K \cdot \ve R)  \cos^2 (k_F z). 
\end{align}
For the susceptibility on different sublattices 
we find
\begin{align}
&\chi^{AB}_0(\ve R,i\omega_n)=-\frac{2a^2}{\pi^2\hbar v_F |z|}\sin^2 ( \ve K \cdot \ve R)\nonumber\\
&\times \left[ \sin^2 (k_F z) I_2 \left(\frac{\omega_n |z|}{v_F}\right) - \cos^2 (k_F z) I_0 \left(\frac{\omega_n |z|}{  v_F}\right)
\right],
\end{align}
with the corresponding dynamical structure factor being given by
\begin{align}
S^{AB}(\ve R, \omega)
=-\frac{2a^2\omega}{\pi^2 v_F^2} \sin^2 ( \ve K \cdot \ve R)  \sin^2 (k_F z). 
\end{align}
The dynamical structure factors $S^{AA}(\ve R,\omega)$ and $S^{AB}(\ve R,\omega)$ are linear in  $\omega$, and, as expected, vanish at zero frequency (we recall that we work at zero temperature).  
Moreover, the noise is strongly suppressed at some special points $R_z$ in space that satisfy  the condition $k_F z = n\pi/2$, where $n$ is an integer.

In the opposite limit  $ {\omega |z|}\gg{ v_F} $ we use the following asymptotics, $I_0(x)\approx i/x+ \pi x e^{-x}/2$ and $I_2(x) \approx -2i /x^3- \pi x e^{-x}/2$ for $x\gg1$. After analytic continuation, the dynamical structure factor is then given by
\begin{align}
S^{AA}(\ve R, \omega)=&\frac{a^2 \omega}{\pi^2 v_F^2}\cos^2 ( \ve K \cdot \ve R)
 \cos \left(\frac{\omega |z|}{  v_F}\right) ,
\\
S^{AB}(\ve R,\omega)=&\frac{a^2\omega}{\pi^2 v_F^2}\sin^2 ( \ve K \cdot \ve R)
\cos \left(\frac{\omega |z|}{  v_F}\right).
\end{align}
In this regime, $S^{AA}(\ve R, \omega)$ and $S^{AB}(\ve R, \omega)$ are not only linearly proportional to the frequency but  also oscillate rapidly as a function of frequency. This implies that in real time the spin noise is only non-zero for times $t$ and distances $z$ satisfying $|z|=v_F t$.

\subsection{Semiconducting nanotubes}

For semiconducting CNTs all calculations for the frequency dependent susceptibility are similar to the ones  for one-dimensional systems with parabolic spectrum, being available in the literature. \cite{giuliani_book} At zero temperature ${\rm Im} \chi_{\alpha \alpha} ({ q},\omega) $ is given by
\begin{align}
&{\rm Im} \chi_{\alpha \alpha} ({q}, \gamma, \gamma', \omega)=-\frac{\pi}{2}\sum_{k}\left[\theta(\xi_{k,\gamma})-\theta(\xi_{k+q,s,\gamma'})\right]\nonumber\\
&\hspace{80pt} \times \delta(\hbar \omega +\xi_{k,s,\gamma}-\xi_{k+q,s,\gamma'}).
\end{align} 
The upper and lower frequencies $\omega_{\pm}$ are defined as
\begin{equation}
\omega_{\pm}(q) = \left|\frac{ \upsilon_F q^2}{2k_G}\pm\frac{\upsilon_F k_F }{k_G} |q|\right|.
\end{equation}
The imaginary part of the spin susceptibility is non-zero only for frequencies 
\begin{equation}
\omega_-(q)\leq|\omega|\leq\omega_+(q) \label{omega_range}
\end{equation}
and is given by
\begin{equation}
{\rm Im} \chi_{\alpha \alpha} ({ q}, \gamma, \gamma', \omega) = -\frac{ak_G}{4\hbar\upsilon_F |q|} {\rm sgn}(\omega).
\end{equation}
To arrive at the expression in real space we perform the Fourier transformation,
\begin{align}
&{\rm Im} \chi_{\alpha \alpha} (\ve R, \omega) = (1\pm\cos (2 \ve K \cdot \ve R))\nonumber\\
&\hspace{50pt} \times\frac{a^2k_G}{4\pi \hbar\upsilon_F}{\rm Im}\int dq\ \frac{e^{i q z}}{|q|}{\rm sgn}(\omega),
\end{align}
where the range of the $q$-integration is determined from Eq.~(\ref{omega_range}), and the positive (negative) sign corresponds to $\chi^{AA}_{\alpha \alpha}$ ($\chi^{AB}_{\alpha \alpha}$). For high frequencies, $\omega>\upsilon_F k_F^2/2k_G$, the dynamical structure factor is given by
\begin{align}
&S(\ve R, \omega)=(1\pm\cos (2 \ve K \cdot \ve R))\frac{a^2k_G}{2\pi^2\upsilon_F} \nonumber\\
&\hspace{40pt}\times\left[{\rm si}(q_+(\omega) |z|)-{\rm si}(q_-(\omega) |z|)\right]{\rm sgn}(\omega),
\end{align}
where wavevectors $q_\pm(\omega)$ are positive solutions of the equations $|\omega|=|\omega_\pm(q_\pm)|$. For low frequencies, $0<\omega\leq\upsilon_F k_F^2/2k_G$, the same equation
$|\omega|=|\omega_-(q_{-,i})|$ has three non-negative solutions, \cite{giuliani_book} $q_{-,1}\leq q_{-,2}\leq  q_{-,3}$. In this case $S(\ve R, \omega)$ is given by
\begin{align}
&S(\ve R, \omega)=(1\pm\cos (2 \ve K \cdot \ve R))\frac{a^2k_G}{2\pi^2\upsilon_F} {\rm sgn}(\omega)\nonumber\\
&\hspace{40pt}\times[{\rm si}(q_+(\omega) |z|)-{\rm si}(q_{-,1}(\omega) |z|)\nonumber\\
&\hspace{60pt}+{\rm si}(q_{-,2}(\omega) |z|)-{\rm si}(q_{-,3}(\omega) |z|)].
\end{align}
We note that the expression is composed of several contributions and thus leads to beating patterns of the spin noise, similar to the one before
for the spin susceptibility.

\section{Graphene nanoribbons \label{sec:ribbons}}

\subsection{The effective Hamiltonian}

In the last part of this work, we turn to graphene nanoribbons, which are finite-size sheets of graphene.\cite{brey_2006} 
The nanoribbon is assumed to be aligned along the $z$-direction and to have a finite width $W=Na$ in $x$-direction, with $N$ being the number of unit cells in this transverse direction. Here, we focus on armchair nanoribbons, characterized by the fact that the $x$-axis points along one of the translation vectors of the graphene lattice. The effective Hamiltonian is given by
\begin{equation}
H_0=\hbar \upsilon_F (\gamma k_x \sigma_1 +  k_z \sigma_2)\, ,
\end{equation}
which determines the low-energy spectrum around the two Dirac points ${\bf K}=-{\bf K}^\prime=(4\pi/3a, 0)$. Here, $k_z$ is the momentum in $z$-direction. The momentum $k_x$ in $x$-direction is quantized due to the vanishing boundary conditions imposed on the extended nanoribbon.\cite{brey_2006} If the width of the GNR is such that $N=3M+~1$, where $M$ is a positive integer,
the GNR is metallic with $k_x=0$. Otherwise, the nanoribbon is semiconducting.

The eigenstates are written as $\psi=\sum_{\sigma\gamma} \phi_{\sigma\gamma} e^{i \gamma K_x x}$, $\Phi=(\phi_{AK},\phi_{BK},\phi_{AK^\prime},\phi_{BK^\prime})$, where $\sigma = A,B$.
 The corresponding  spectrum and wavefunctions that satisfy the vanishing boundary conditions (for $\psi$) are given by 
\begin{align}
&\Phi^{\epsilon,k_z}_{\zeta} =e^{i k_z z } (   -i \zeta, 1, i \zeta , -1),
\label{met_wavefunc}\\
&\epsilon_{\zeta}= \zeta \hbar \upsilon_F k_z \label{metal_spectrum}
\end{align}
for a metallic GNR, and
\begin{align}
&\Phi^{\epsilon,k_z}_{\pm} =e^{i k_z z } (\pm e^{i\varphi_s+i xk_x^{min} }, e^{i xk_x^{min} },\nonumber\\ 
&\ \ \ \ \ \ \ \ \ \ \   \ \ \ \ \ \ \ \ \ \ \ \  \ \  \mp e^{i\varphi_s-i xk_x^{min} }, 
 -e^{-i xk_x^{min} }),
 \label{sem_wavefunc}\\
&\epsilon_{\pm}= \pm \hbar \upsilon_F \sqrt{ (k_x^{min})^2+k_z^2}
\end{align}
for a semiconducting GNR. Here, $\zeta=\pm 1$ is the eigenvalue of the Pauli matrix $\sigma_2$, and we use the notation $e^{i\varphi_s}=[k_x^{min} - i k_z]/\sqrt{(k_x^{min})^2 +k_z^2 }$,  with $|k_x^{min}|=\pi/3(N+2)a$.

\subsection{Spin susceptibility}
 
\subsubsection{Without SOI}

To calculate the spin susceptibility for a metallic nanoribbon that has a linear spectrum given by Eq. (\ref{metal_spectrum}), we again work in the imaginary time formalism, see Eq. (\ref{kogan_chi}). The calculations are quite similar to the ones presented before in Sec. \ref{Sec:matallic}. The only change in the expressions for the spin susceptibility in comparison with a metallic nanotube [see Eqs. (\ref{met_aa}) and (\ref{susc_metallic_cnt})]  is in the fast oscillating prefactor,
\begin{align}
\chi_0^{AA}(\ve R_i,& \ve R_j)=-\chi_0^{AB}(\ve R_i, \ve R_j)=-\frac{a^2}{2\pi \hbar v_F |z|}\nonumber\\
&\times\sin^2(\ve K\cdot \ve R_i)
\sin^2(\ve K\cdot \ve R_j)\cos (2 k_F z)\label{aa_rib}.
\end{align}
Similarly, for semiconducting nanoribbons the spin susceptibility is given by Eqs. (\ref{semi_AA}) and (\ref{semi_AB}), where the fast oscillating prefactors $1\pm\cos (2 \ve K \cdot \ve R)$ are replaced by $\sin^2(\ve K\cdot \ve R_i)\sin^2(\ve K\cdot \ve R_j)$.

\subsubsection{With SOI}

The intrinsic SOI in graphene is only several $\mu eV$, so it is rather weak. Moreover, the Rashba SOI generated by an externally applied electric field $E$ is in the range of tenths of $\mu eV$ for $E=1\ \rm{V/nm}$. \cite{cnt_helical_2011,klinovaja_MF_CNT}
Such small SOI values might be hard to observe. However, the Rashba SOI generated by a spatially varying magnetic field opens new perspectives for spintronics in graphene. \cite{klinovaja_Nanoribbon} In this case, the SOI strength can be exceptionally large, reaching hundreds of $\rm {meV}$. A nanoribbon in the presence of a rotating magnetic field with period  $2\pi/k_n$, described by the Zeeman Hamiltonian
\begin{equation}
H_n^\perp=\Delta_Z\left[ s_y \cos (k_n z)+s_z \sin (k_n z)\right],
\end{equation}
is equivalent to a nanoribbon with Rashba SOI in the presence of  a uniform magnetic field,
\begin{align}
H^\perp&=U_n^\dagger (H_0+H_n^\perp) U_n = H_0 + \Delta_Z s_y + \Delta_{so}^n s_x \sigma_2,
\end{align}
where the unitary gauge transformation is given by $U_n= \exp(i k_n z s_x/2)$.
The period of the magnetic field determines the strength of the Rashba SOI $\Delta_{so}^n=\hbar \upsilon_F k_n/2$, while the
amplitudes of the uniform and the rotating fields are the same and given by $\Delta_Z $.\cite{Braunecker_Jap_Klin_2009, klinovaja_Nanoribbon} 
The spectrum of a metallic (semiconducting) GNR in the presence of such SOI consists of two cones (parabolas) shifted along the momentum axis against each other  by $k_n/2$. Every branch of the spectrum
possesses a well-defined spin polarization perpendicular to the $z$-axis that is along the nanoribbon. 
A uniform magnetic field only slightly modifies the spectrum by opening a gap at zero momentum. In the following discussion we neglect the uniform magnetic field working in the regime where the induced Rashba SOI is stronger than the Zeeman energy $2\Delta_Z $.

As a result, similar to the semiconducting nanowire, one can gauge away the momentum shifts by rotating the spin coordinate system as follows,
\begin{align}
&I_x (\ve R) = I_x \cos (k_n z)  +I_y \sin (k_n z) ,\\
&I_y (\ve R)= I_y \cos (k_n z)  - I_x \sin (k_n z) ,\\
&I_z (\ve R) =I_z.
\end{align}
The same transformation should be applied to the electron spin operators $\hbar {\bf s}/2$.
The effective RKKY Hamiltonian in this rotated coordinate system is the same as in the system without SOI and is given by Eq. (\ref{aa_rib}). To return to the laboratory frame, we perform the following change
\begin{align}
\ve I_i \cdot \ve I_j&\rightarrow \ve I_i ( \ve R_i) \cdot \ve I_j ( \ve R_{j})= \cos (k_n z) \ve I_i \cdot \ve I_j\nonumber\\
& + [1-\cos (k_n z)] I^z_i  I^z_j -  \sin (k_n z)(\ve I_i \times \ve I_j)_z.
\end{align}
The spin susceptibility tensor has non-vanishing off-diagonal components, which, again, indicate a broken invariance of spin space
induced by the magnetic field or the Rashba SOI. As before, this gives rise to anisotropic RKKY interactions of Ising and Moryia-Dzyaloshinski form.

\section{Conclusions \label{sec:concl}}

In the present work we studied the Rudermann-Kittel-Kasuya-Yosida (RKKY) interaction in carbon nanotubes and graphene nanoribbons at zero temperature 
in the presence of spin orbit interaction.
Our main results are summarized in the following. 

The spin susceptibility in metallic CNTs, characterized by a Dirac spectrum (gapless and linear), crucially depends on whether the localized spins that interact with each other are from the same or from different sublattices.
In particular, if the Fermi level is tuned exactly to the Dirac point where the chemical potential is zero the interaction is of ferromagnetic type for spins on $A$-$A$ or $B$-$B$ lattice sites, whereas it is of  antiferromagnetic type for spins on $A$-$B$ lattice sites.  
In semiconducting CNTs, with a sizable bandgap, the spin susceptibility depends only slightly on the sublattices. In all cases, the spin susceptibility is an oscillating function that decays as $1/R$, where $R$ is the distance between the localized spins. 

The spin orbit interaction breaks the spin degeneracy of the spectrum and the direction invariance of the spin space. As a result, the spin susceptibility is described by the tensor $\chi_{\alpha \beta}$ that has two non-zero off-diagonal components $\chi_{xy}=-\chi_{yx}$, the finite values of which signal the presence of SOI in the system. Moreover, the RKKY interaction is also anisotropic in the diagonal terms, $\chi_{zz}\neq\chi_{xx}=\chi_{yy}$. Quite surprisingly, we find that all non-zero components, diagonal and off-diagonal, can be tuned to be of equal strength by adjusting the Fermi level.
These anisotropies, giving rise to Ising and Moriya-Dzyaloshinski RKKY interactions, thus open the possibility to have magnetic order in  low-dimensional systems at finite temperature. \cite{order_2_2011}

We note that, in contrast to semiconducting nanowires, the SOI  cannot be gauged away by a unitary transformation  in CNTs, giving rise to a more complicated dependence of $\chi_{\alpha\beta}$ on the SOI parameters. In the same way, a magnetic field along the CNT axis breaks both the spin and the valley degeneracy, leading to a dependence of the spin susceptibility on four different Fermi wavevectors. 

The spin susceptibility at finite frequencies also allows us to analyze the spin noise in the system via the fluctuation-dissipation theorem. We find that the dynamical structure factor $S_{\alpha\alpha}(\ve R, \omega)$ is linear in frequency and oscillates in real space.

Metallic armchair GNRs behave similarly to metallic CNTs. Indeed, in both cases the spin susceptibility  shows a strong dependence on the sublattices, with, however, different fast oscillating prefactors. A Rashba-like SOI interaction  can be generated in armchair GNR by periodic magnetic fields. In contrast to  CNTs with intrinsic SOI, this field-generated SOI can be gauged away giving rise to a simple structure of the spin susceptibility tensor.\cite{note_bilayer}

In this work we have ignored interaction effects. However, it is well-known that in one- and two-dimensional systems electron-electron interactions can lead to interesting  modifications of the spin susceptibility, for instance with a slower power law decay such as $1/R^g$, with $0<g\leq 1$ in a Luttinger liquid approach
to interacting one-dimensional wires. \cite{Egger_1996,Braunecker_2009_PRL} It would be interesting to extend the present analysis  and to allow for interaction effects\cite{Imambekov_RMF} in the spin susceptibility for carbon based materials in the presence of spin orbit interaction, in particular for metallic CNTs and GNRs at the Dirac point.

\acknowledgments
This work is supported by the Swiss NSF, NCCR Nanoscience, and NCCR QSIT.


\begin{thebibliography}{100}
\bibitem{note_bilayer}  We note that one-dimensional quantum wires can also be formed in gated bilayer graphene.\cite{martin_bilayer} In this system, an effective SOI can be created by a fold (curvature) or by rotating magnetic fields.\cite{klinovaja_bilayer} The RKKY interaction in such quantum wires has a form similar to the one considered in Sec. \ref{sec:ribbons}.
\end{thebibliography}
\end{document}